\documentclass[final,3p,times]{elsarticle}
\biboptions{sort&compress}

\usepackage{hyperref}

\usepackage{graphicx}

\usepackage{amssymb}
\usepackage{amsmath}
\usepackage{upgreek}
\usepackage{lineno}
\usepackage{gensymb}
\usepackage[utf8]{inputenc}
\usepackage[dvipsnames]{xcolor}

\usepackage{fancyhdr}
\usepackage[normalem]{ulem}

\journal{a journal}

\begin{document}

\begin{frontmatter}

\title{Mechanical and electrical properties of a nano-gap or how to play the nano-accordion}

\author[label1,label2,label3]{Simon Hettler\corref{cor1}}
\author[label2,label3,label4]{Raul Arenal}
\address[label1]{Laboratory for Electron Microscopy, Karlsruhe Institute of Technology, Karsruhe, Germany}
\address[label2]{Instituto de Nanociencia y Materiales de Aragón (INMA), CSIC-Universidad de Zaragoza, Zaragoza, Spain}
\address[label3]{Laboratorio de Microscopías Avanzadas (LMA), Universidad de Zaragoza, Zaragoza, Spain}
\address[label4]{ARAID Foundation, Zaragoza, Spain}
\cortext[cor1]{simon.hettler@kit.edu}

\begin{abstract}

\textit{In-situ} transmission electron microscopy (TEM) has become an important technique to study dynamic processes at highest spatial resolution and one branch is the investigation of phenomena related with electrical currents. Here, we present experimental results obtained from a peculiar \textit{in-situ} TEM device, which was prepared with the aim to analyze the relationship between (thermo)electric properties and specific crystal orientations of a misfit layered compound. The formation of a nano-sized gap at a grain boundary facilitated a precisely controllable mechanical bending of the device by application of differential heating currents. The devices' electrical properties were found to be substantially influenced by the gap, leading to a high intrinsic voltage. This voltage additionally depends on the vacuum environment and on the history of applied heating currents. These findings are largely attributed to the presence of adsorbed molecules within the gap region. The electrical \textit{in-situ} TEM studies of this work illustrate that interior surfaces can strongly influence electrical properties even under high vacuum conditions.

\end{abstract}

\begin{keyword}
\textit{in-situ} transmission electron microscopy \sep nano-gap \sep grain boundary \sep mechanical bending \sep intrinsic voltage
\end{keyword}
\end{frontmatter}

\section{Introduction}
\label{S:Intro}

Electrical conduction in matter is mainly governed by the materials' band structure and electron scattering processes in the material. In pristine single-crystalline bulk specimens and at finite temperatures and bias, electron-phonon interaction is the limiting factor for the conductivity. In real samples, scattering from defects in the material constitutes an additional contribution. The microstructure, including interfaces, surfaces, grain boundaries and point defects, can therefore significantly alter the electrical conductivity of real materials in comparison to defect-free single crystals. The relationship between microstructure and electrical conductivity has been subject of investigation for a long time \cite{Mayadas1970,Andrews1969} and research is continuing until today, with the motivation to improve the conductivity of electrical interconnects or the performance of thermoelectric materials \cite{Fava2021,MYPATI2022100223,BUENOVILLORO2023118816,DONG20231459}. 

Due to the continuing miniaturization of electronic devices, microscopic approaches to analyze the impact of individual defects on the electrical conductivities have been followed in the past decades \cite{Graham2010,Bishara2021}. Specifically, \textit{in-situ} TEM has received interest due to its high spatial resolution \cite{Aslam2011,Hettler_2021,Hsueh2023,Hettler2024}. Only recently, the possibility of performing \textit{in-situ} TEM thermoelectric characterizations has been demonstrated \cite{Hettler2025}. Several Misfit-layered compound (MLC) families are promising materials for thermoelectric applications \cite{Ng2022}. In bulk MLC materials, both grain boundaries and pore density are expected to have a strong impact on their performance. 

This work is part of an experiment series, where different \textit{in-situ} TEM specimens of a strontium-doped variant of the calcium cobaltite (CCO) MLC family (Ca\textsubscript{2.93}Sr\textsubscript{0.07}Co\textsubscript{4}O\textsubscript{9}) \cite{Torres.2022} have been prepared with the aim of elucidating the impact of grain boundaries and crystallographic directions on the (thermo)electric properties. Here, we present experimental results obtained from one peculiar \textit{in-situ} TEM device of this series, which exhibited a nano-sized gap and will be named nano-accordion throughout the manuscript. The name is motivated by the controlled mechanical movement of the device induced by an applied differential heating current, which resembles the folding and unfolding of an accordions' bellows. The gap also has a strong impact on the electrical characteristics, manifesting itself as a large intrinsic voltage offset, which is attributed to adsorbed molecules present in the gap region.

\section{Materials and Methods}
\label{S:MaM}

\subsection{\textit{In-situ} TEM}

\textit{In-situ} transmission electron microscopy (TEM) investigations were carried out using two aberration-corrected (one probe-, the other image-corrected) Titan microscopes (Thermo Fisher Scientific) operated at 300~keV. Both microscopes were used for bright-field (BF) and dark-field (DF)-TEM imaging as well as for selected-area electron diffraction (SAED). High-angle annular dark field (HAADF) scanning (S)TEM was performed in the probe-corrected microscope and high-resolution (HR)TEM imaging was performed in the image-corrected microscope. SAED patterns were simulated using the ReciPro software \cite{Seto2022}.

\begin{figure}[t]
\centering
    \includegraphics[width=0.65\linewidth]{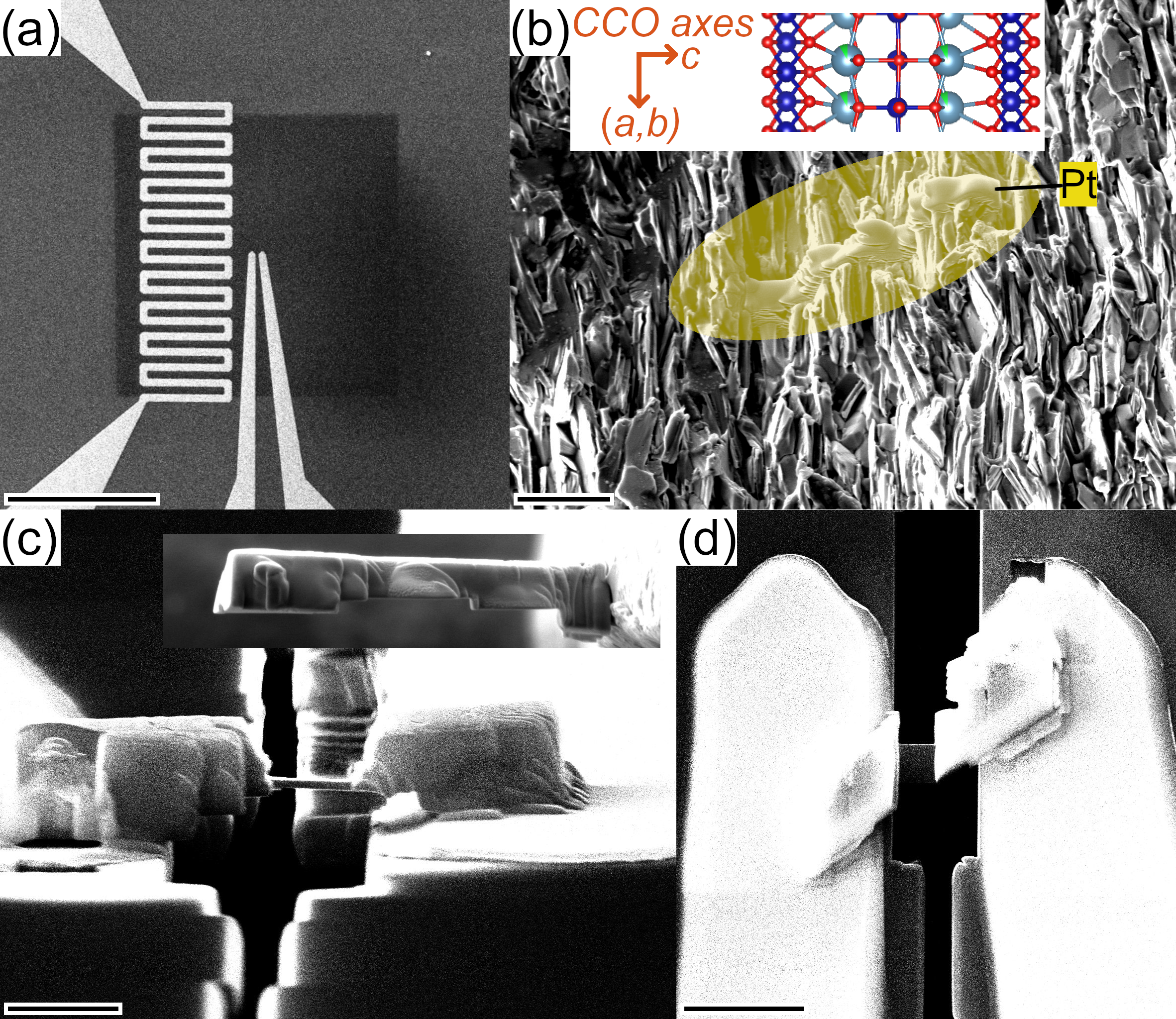}
    \caption{Preparation of the nano-accordion. (a) SEM image of the membrane area with the differential heating element and the contact pads of the \textit{in-situ} chip, see (d) for a zoomed-in view of the pads. (b) SEM image of the CCO material reveals its sheet-like structure, the predominant \textit{c} axis direction of the material has been indicated. A sketch of the MLC structure with Co (dark blue), Ca/Sr (light blue/green) and O (red) atoms has been added. The deposited Pt, highlighted in yellow, indicates the place from where the lamella was taken and its orientation with respect to the CCO sheets. Stage tilt is 52\degree. (c) SEM image of the thinned device on the \textit{in-situ} chip with an incident angle of the electron beam of 5\degree with respect to the surface of the chip. The two Pt contact pads, the thick and thin parts of the lamella as well as the long hole in the membrane are visible. The inset shows the bridge-shaped form of the lamella after transfer to a Cu TEM grid. (d) Top view of the final \textit{in-situ} device.  Scale bars are (a) 200~$\upmu$m, (b) 10~$\upmu$m, (c) 2~$\upmu$m, width of inset SEM image is 15~$\upmu$m and (d) 5~$\upmu$m.}
    \label{F:Figure1}
\end{figure}

A DENSolutions Wildifre specimen holder with a custom chip designed for \textit{in-situ} thermoelectrical characterizations \cite{Hettler2025} was used to conduct the \textit{in-situ} TEM experiments. Fabrication of the chip is described elsewhere \cite{Hettler2025}. The chip contains a silicon nitride membrane with a thickness of 1~$\upmu$m in its center, on which two contact pads and a differential heating element have been structured by lithography. A scanning electron microscopy (SEM) image of the membrane area of the chip is shown in Figure~\ref{F:Figure1}a. The application of a heating current to the differential heating element results in the generation of a temperature gradient along a specimen that is placed between the two contact pads. A long hole of more than 50~$\upmu$m length is milled in the silicon nitride membrane between the two pads, which acts as thermal insulation barrier and thus ensures the generation of a large temperature gradient \cite{Hettler2025}. The hole is (partially) seen in the SEM images of the final device shown in Figures~\ref{F:Figure1}c and d.

\subsection{Specimen preparation}

The \textit{in-situ} TEM specimen was prepared using a Helios 650 dual-beam instrument (Thermo Fisher Scientific) combining SEM with a Ga\textsuperscript{+} focused ion beam (FIB). In a first step, a conventional TEM lamella was prepared from a strontium-doped calcium cobaltite (CCO: Ca\textsubscript{2.93}Sr\textsubscript{0.07}Co\textsubscript{4}O\textsubscript{9}) misfit-layered compound (MLC). The preparation of the bulk CCO material is described elsewhere \cite{Torres.2022}. The CCO material consists of the two subsystems CoO\textsubscript{2} and Ca\textsubscript{2-x}Sr\textsubscript{x}CoO\textsubscript{3}, which are stacked alternately along the \textit{c} axis of the incommensurate crystal structure, which is sketched in the inset of Figure~\ref{F:Figure1}b. Figure~\ref{F:Figure1}b also shows an SEM image of the CCO material, which exhibits numerous sheets in vertical direction of the image. The out-of-plane direction of the sheets (horizontal in the image) roughly corresponds to the \textit{c} axis of the CCO material. The TEM lamella was prepared with the \textit{c} axis of the CCO material coinciding with the long axis of the lamella, which is seen from the deposited stripe of Pt in Figure~\ref{F:Figure1}b, which served as protection layer for the lamella. The lamella was then transferred to a TEM grid using a standard lift-out method. The lower side of the lamella was polished to guarantee a good contact with the \textit{in-situ} chip in the next step. In the central part, several hundreds of nanometers of the lamella were removed on the lower side, resulting in a bridge-like shape. This bridge-like shape facilitates the later thinning of the lamella on the \textit{in-situ} chip and is revealed in the inset image in Figure~\ref{F:Figure1}c. The inset image shows the top-view of the lamella on the grid. 

To transfer the lamella to the \textit{in-situ} chip, both the chip and the grid (lying flat on the surface) were mounted on separate stubs in the dual-beam instrument. First, a long hole was milled between the contact pads of the \textit{in-situ} chip by FIB. The lamella was then lifted out from the TEM grid, deposited on the chip, and fixed by FIB-induced deposition of Pt. For thinning, the chip was mounted on a stub with a surface with a vertical inclination of 45\degree. Thinning of the lamella was performed at a stage tilt of 12\degree, resulting in a FIB incident angle on the surface of the chip of 5\degree~ (Figure~\ref{F:Figure1}c) due to the angle of 52\degree between FIB and electron beam. After thinning, the Pt protection cover was removed from the thin part of the device by FIB milling, leaving the thinned  part with an approximately quadratic shape with a side length of 1.6~$\upmu$m and a thickness of 120~nm (Figure~\ref{F:Figure1}c). FIB operation voltage was 30~kV for all steps and final thinning current was 24 pA. 

\subsection{Electrical characterization}

Two Keithley Instruments (Tektronix) devices were used for electrical characterization of the specimen (2450 SourceMeter) and simultaneous application of a heating current $I_H$ to the differential heating element on the chip  (2611A System Source Meter). The heating current was set manually, and the I-V characterization was performed with the Keithley Kickstart software.

\newpage

\section{Results and discussion}

The CCO device was prepared with the aim of studying its thermoelectric properties, but the electrical measurements showed unexpected characteristics. A detailed TEM analysis revealed that a nano-gap formed in the device, which is described first in section~\ref{S:3.1}. The mechanical properties of the device are described in the following section~\ref{S:3.2} before the electrical characterization is presented and discussed in section~\ref{S:3.3}. Finally, the device characterization after the formation of a fully penetrating gap is presented in section~\ref{S:3.4}.

\subsection{Formation of a nano-gap at a grain boundary}
\label{S:3.1}

\begin{figure}[t]
\centering
    \includegraphics[width=0.9\linewidth]{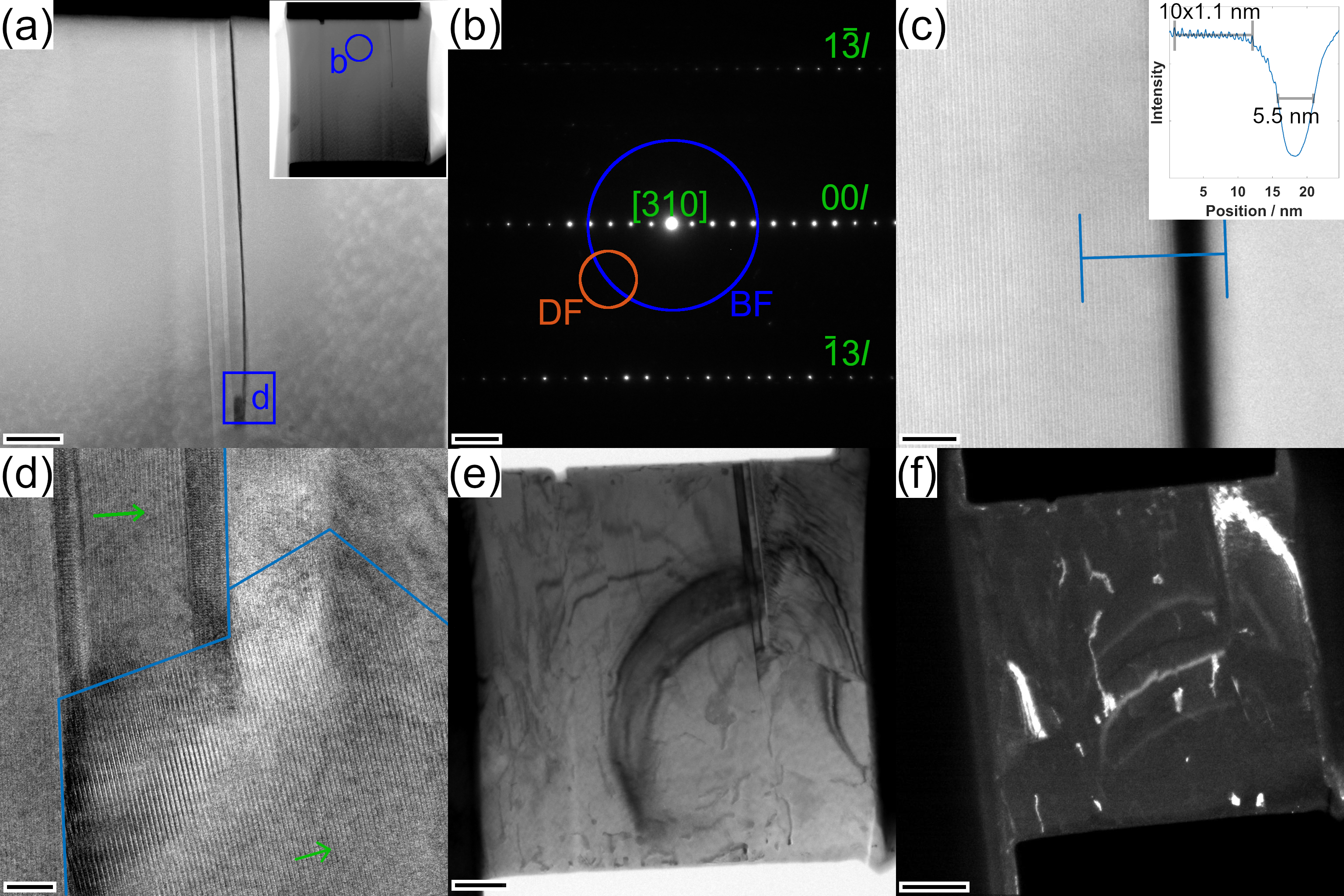}
    \caption{(a) HAADF-STEM image of the nano-accordion at a stage tilt of 11.5\degree ~revealing the presence of a gap in the device. The inset STEM image shows the entire region of the thin lamella. The selected area for the SAED pattern shown in (b) and the acquisition position of the HRTEM image shown in (d) are indicated. (b) SAED pattern taken from the area marked in (a) revealing the horizontal orientation of the \textit{c} axis of the material in the left part of the lamella. The sets of [00\textit{l}] and [13\textit{l}] reflections are named and the placement position of the objective aperture for acquisition of the BF and DF-TEM images shown in (e,f) is indicated. (c) HRSTEM image of the central gap region revealing the stack along the \textit{c}-axis on its left side. A line profile taken from the marked area is shown in the inset revealing a stack periodicity of 1.1~nm and a gap width of approximately 5.5~nm. (d)~HRTEM image of the end of the gap with grain boundary positions (blue lines) and the direction of the respective \textit{c}-axis orientation (green arrows) marked. (e) BF- and (f) DF-TEM images of the thin part of the lamella, description see text. Scale bars are (a) 100~nm, inset image width is 2~$\upmu$m, (b) 2~nm\textsuperscript{-1}. (c) 9~nm, (d) 10~nm, (e) 200~nm and (f) 300~nm.}
    \label{F:Figure2}
\end{figure}

Figure~\ref{F:Figure2}a shows a HAADF-STEM image of the thin part of the device at a stage tilt of 11.5\degree. At this tilt, the image reveals a dark vertical line, which corresponds to a nano-sized gap separating the left from the right sections of the lamella. The inset image shows that the gap extends over approximately half of the height of the lamella. An SAED pattern taken from the left side of the gap (Figure~\ref{F:Figure2}b) shows that in this region, the \textit{c} axis of the CCO material agrees with the horizontal direction. The orientation of the CCO crystal is close to the [310] orientation of the Ca\textsubscript{2-x}Sr\textsubscript{x}CoO\textsubscript{3} subsystem as both sets of [00\textit{l}] and [13\textit{l}] reflections can be identified in the SAED pattern (Figure~\ref{F:Figure2}b). In contrast, the electron beam direction does not agree with a higher-order crystal direction of the CCO material in the right part as revealed by a corresponding SAED pattern, shown in the supplementary information, Figure~S1. 

The gap thus formed at a grain boundary of the material, which is also seen from high-resolution (HR) STEM and TEM images taken from the gap region (Figures~\ref{F:Figure2}c and d, respectively). The HRSTEM image shows the stack periodicity on the left side of the gap. A line scan across the border allows to measure a periodicity of 1.1~nm (inset image in Figure~\ref{F:Figure2}c), which agrees with the reference value of 1.08~nm of the CCO material \cite{Miyazaki2002}. The line scan also reveals a gap width at this position of approximately 5.5~nm. The HRTEM image shown in Figure~\ref{F:Figure2}d was acquired from the lower end of the gap and reveals that the gap originates at a triple grain boundary. Three grains are visible in the image and their borders have been marked by blue lines. The grain on the upper left side shows a horizontal \textit{c}-axis orientation, indicated by a green arrow in Figure~\ref{F:Figure2}d, which is in agreement with the SAED pattern and the HRSTEM image in Figure~\ref{F:Figure2}b and c. In the lower grain, the stack periodicity is visible, but with a rotation compared to the previous grain. Finally, on the upper right side, no lattice fringes can be detected. At this position, the gap is not penetrating through the entire thickness of the lamella, but the partial gap is visible from the area with bright contrast forming along the grain boundary of the upper left grain.

Figures~\ref{F:Figure2}e and f show a bright-field (BF)-TEM and dark-field (DF)-TEM image of the lamella region with the respective placement of the objective apertures indicated in Figure~\ref{F:Figure2}b. In the BF-TEM image, dark areas correspond to regions, where an increased number of electrons undergo diffraction or scattering events to higher angles than given by the objective aperture radius. As the lamella is made of a single material and exhibits a homogenoeus thickness, the visible features correspond to diffraction contrast. A dark bend stripe can be seen in the image that originates in the lower part of the lamella and ends at the gap, indicating that in this area, the crystal exhibits a higher-order orientation with respect to the electron beam. At the same time, it indicates that the lamella is strongly bend as a relaxed grain with similar orientation throughout would exhibit a homogeneous contrast. The same observation can be made from the DF-TEM image shown in Figure~\ref{F:Figure2}f. The objective aperture was placed in a region where no major reflections of the [310] orientation is located (see orange circle in Figure~\ref{F:Figure2}b). Several small bright areas are visible in Figure~\ref{F:Figure2}f that are caused by inclusions with different crystal orientation. In addition, a bright stripe is visible in the upper right grain, which indicates that also the right grain is strongly bend. Three thin white lines can be seen in the left grain, that are assumed to correspond to an additional minor reflection, possibly the [020] reflection of the Ca\textsubscript{2-x}Sr\textsubscript{x}CoO\textsubscript{3} subsystem. 

This analysis of the nano-accordion reveals that there is a gap in the device and that the material is considerably bend. The formation of the gap could be attributed to two effects: Firstly, pores are frequently encountered in the CCO material and the gap could have corresponded to an elongated pore. Secondly, another plausible explanation is the opening of the gap at the grain boundary due to mechanical stress. On the one hand, the high-pressure fabrication process introduces considerable stress in the bulk material. The opening of the gap could be caused by a release of that stress. On the other hand, the lamella is subjected to possible mechanical stress on the device induced by the SiN membrane after cutting of the hole.

\subsection{Thermally induced controlled bending of the specimen}
\label{S:3.2}

\begin{figure}[t]
\centering
    \includegraphics[width=0.9\linewidth]{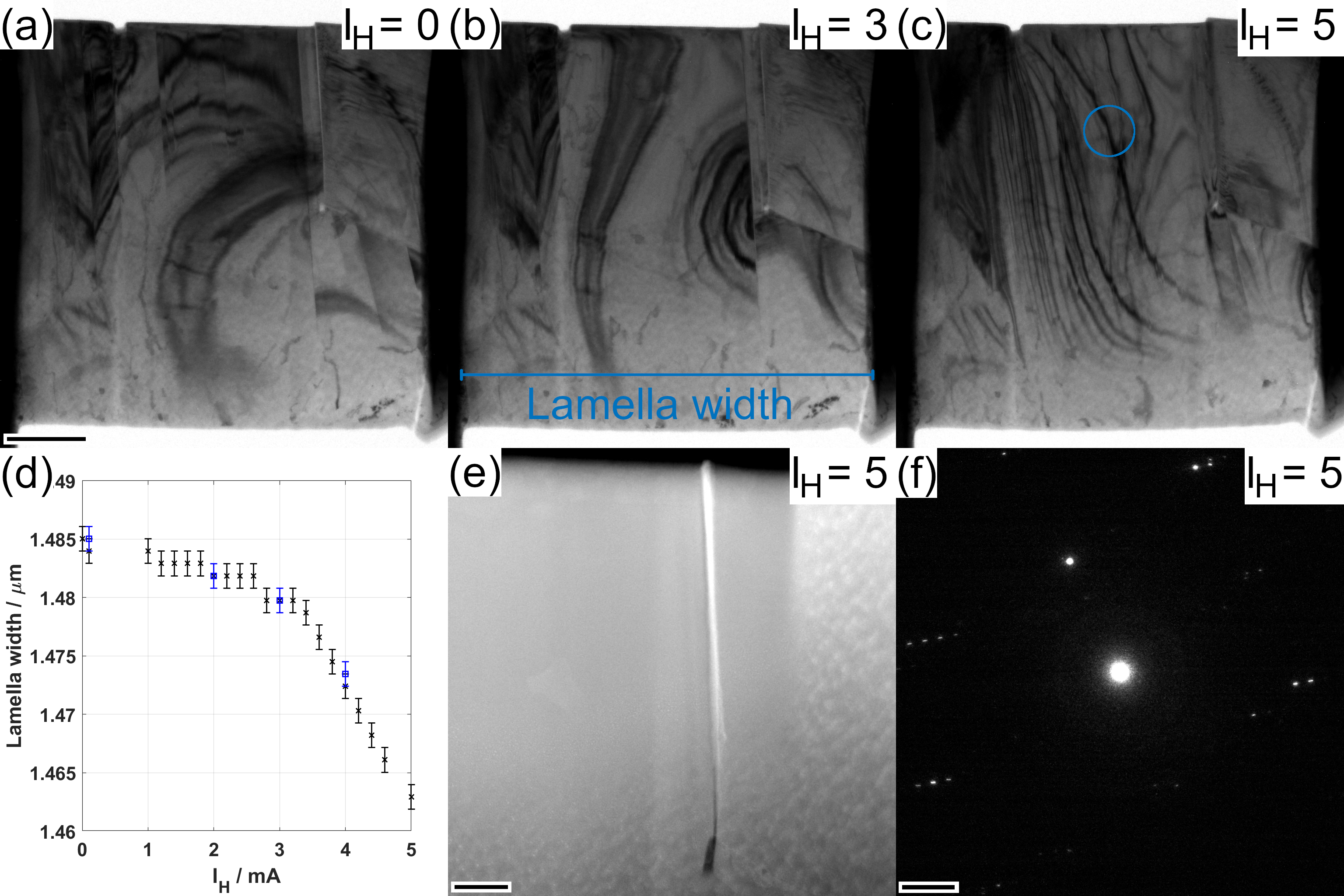}
    \caption{(a-c) BF-TEM images of the thin part of the device for different applied $I_H$ of (a) 0~mA, (b) 3~mA and (c) 5~mA. The movement of the bending contours due to temperature-induced bending is clearly visible. (d) Plot of the measured width of the lamella as indicated by the blue line in (b) over $I_H$. The non-linearly decreasing width due to thermal expansion and bending is clearly visible. Black crosses correspond to the rise of the current and blue boxes to the descent. Error bars correspond to the pixel size. (e) HAADF-STEM image of the device with $I_H$~=~5~mA. The two parts of the lamella overlap in the upper part due to temperature-induced bending, which leads to the increased contrast in the overlapping region. (f) SAED pattern acquired from the area marked by a blue circle in (c) for $I_H$~=~5~mA. Due to bending, the electron-beam direction is now far from the initial proximity to the [310] orientation. Scale bars are (a-c) 300~nm. (e) 100~nm and (f) 3~nm\textsuperscript{-1}.}
    \label{F:Figure3}
\end{figure}

In this section, the mechanical deformation of the device in response to the heating current $I_H$ applied on the differential heating element is analyzed. For this analysis, the evolution of the device during the application of increasing $I_H$ up to a maximum of 5~mA was followed by different TEM imaging and diffraction techniques. Figure~\ref{F:Figure3}a-c shows three BF-TEM images acquired at $I_H$~=~0, 3 and 5~mA. The whole series of BF-TEM images is given in supplementary video S1. In the initial, unheated state (Figure~\ref{F:Figure3}a), the dark broad bending contour corresponds to the [310] crystal orientation, as discussed previously for Figure~\ref{F:Figure2}e. With increasing applied heating current, the device starts to bend and the bending contour moves towards the left border of the lamella. From the gap region, additional bending contours appear, which can be linked to another minor zone-axis (Figure~\ref{F:Figure3}b). At maximum $I_H$, this additional contours have similarly traveled towards the left border of the lamella. A movement of the contours in the opposite direction and with reduced speed is visible in the two grains located at the right border of the lamella, suggesting a reduced strength of the bending in that region. A difference is also seen between the top part, which includes the gap, and the bottom part, where the material is not separated by a gap. The movement of the contours is enhanced in the top part due to the presence of the gap.

The BF-TEM images also allow to measure the distance between the edges of the thick portion of the lamella as indicated by the blue line in Figure~\ref{F:Figure3}b. The measured width is plotted over $I_H$ in Figure~\ref{F:Figure3}d and reveals a non-linear decrease, which reaches its minimum at $I_H$~=~5~mA. The width decreases by 22~nm when compared to the unheated state. The observed bending and the decreasing width are caused by a thermal expansion of the heated area of the \textit{in-situ} chip. The close to quadratic dependence of the width is explained by the fact that the dissipated heat increases approximately with $I_H^2$. 




In addition to BF-TEM imaging, the device was studied by DF-TEM imaging with similar conditions as in Figure~\ref{F:Figure2}f and the corresponding images can be seen in supplementary video S2. The video shows a similar movement of the bending contours, confirming the considerable bending induced by the applied heating current $I_H$. Supplementary video S3 shows the evolution of the HAADF-STEM contrast in dependence of $I_H$. The HAADF-STEM image of the series acquired at $I_H$~=~5~mA is displayed in Figure~\ref{F:Figure3}e. The image shows the gap-region of the device and instead of the dark line corresponding to the gap, a bright line with increasing width toward the upper border can be seen. This bright line is explained by the overlap between the right and left part of the lamella caused by the bending of both sides toward each other. Supplementary video S3 also reveals the change in channeling contrast due to bending and the related loss of parallelity of the electron beam with the a-b plane of the crystal close to the gap.

The SAED patterns for increasing $I_H$ up to 5~mA can be seen in supplementary video S4 and Figure~\ref{F:Figure3}f shows the SAED pattern at the maximum current of 5~mA. The pattern reveals that at this current, the electron beam current direction does not agree with a crystal plane of the CCO material. The video reveals a smooth evolution of the SAED pattern. The experimental SAED patterns were compared qualitatively with simulated patterns considering only the Ca\textsubscript{2-x}Sr\textsubscript{x}CoO\textsubscript{3} subsystem of the CCO material (see supporting information, Figure~S2). The comparison reveals that the orientation of the crystal with respect to the electron beam changes by approximately 8\degree
~between the unheated state and $I_H$~=~5~mA.

The presented analysis clearly shows that an applied heating current allows to control the mechanical state of the device. This control is highly reproducible, a repeated applied heating current results in the identical mechanical state of the device as shown by the comparison of TEM images and SAED patterns for the same $I_H$ at different points of time (see supporting information, Figure S3). The controlled movement of the device resembles the movement of the bellows during the operation of an accordion (musical instrument) and for that reason, the device was named nano-accordion.

\subsection{Electrical characterization of the nano-gap}
\label{S:3.3}

\begin{figure}[t]
\centering
    \includegraphics[width=0.9\linewidth]{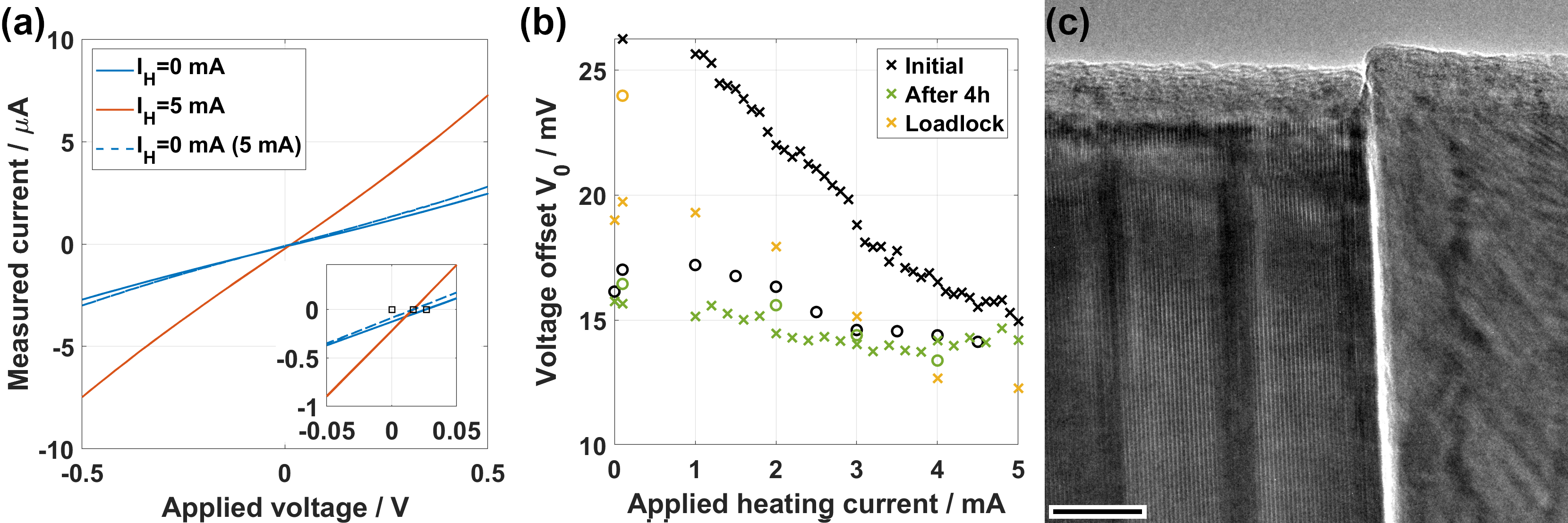}
    \caption{(a) Comparison of three I-V curves for $I_H$~=~0~mA (blue solid line), $I_H$~=~5~mA (red line) and again at 0~mA after having heated with 5~mA (blue dashed line). The inset shows the curves at low voltages and the black squares mark the origin and the voltages where blue and red curves pass zero measured current. (b) Plot of the voltage offset (as marked by black squares in the inset in (a)) for three different conditions of the device: Directly after insertion in the microscope (black crosses and circles), after 4h in the microscope (green) and after retraction to the loadlock (yellow). (c) TEM image of the gap region directly after insertion showing a bright contrast in the gap region. Scale bar is 20~nm.}
    \label{F:Figure4}
\end{figure}

The employed \textit{in-situ} chip allows to study the electrical characteristics of the device in dependence of the applied differential heating current $I_H$. Figure~\ref{F:Figure4}a shows three I-V curves obtained 1) at $I_H$~=~0~mA shortly after insertion of the device in the microscope (solid blue line), 2) at $I_H$~=~5~mA (solid red line) and 3) again in the unheated state after turning of the heating current (blue dashed line). The measured current is considerably increased in the heated case, which can be attributed to the semiconducting nature of the CCO material, whose conductivity increases with the temperature. The current has a non-linear dependency on the applied voltage, which is again explained by the semiconductivity of the material in combination with the considerable current densities ($>$1.5$\cdot$10\textsuperscript{3} Acm\textsuperscript{-2}) at $\pm$0.5~V and an expected Joule heating of the device. 

The chip was designed such that an applied $I_H$ leads to a temperature gradient along the device, which permits the study of the thermoelectric properties of the material \cite{Hettler2025}. To measure that induced voltage, the voltage at zero current is determined in dependence of the applied heating current. Three black squares are added to the inset in Figure~\ref{F:Figure4}a, which indicate the origin (left square) and the voltage offset for the red and blue solid curves (other squares). It is seen that even at no applied heating current, a strong voltage offset of 25~mV is measured, being too high to be a measurement error, which is typically well below 0.1~mV. The voltage offset is decreased for $I_H$~=~5~mA and, more interestingly, the second I-V curve taken at 0~mA after heating, does not agree with the initial measurement.

Figure~\ref{F:Figure4}a shows only three curves of a measurement series for varying $I_H$. For each I-V curve, the voltage offset was determined and is plotted in black crosses and circles over $I_H$ in Figure~\ref{F:Figure4}b. The crosses correspond to the rise and the circles to the descent of the current and the difference is clearly seen. The same measurement series was repeated after 4h of the device in the microscope (green crosses and circles), which does not show a significant difference between rise and descent and which agrees well with the descent of the initial measurement series. These observations suggest that the device undergoes a structural change during the first heating and remains in that state even if a current is applied again. In addition, Figure~\ref{F:Figure4}b shows yellow crosses and circles, which correspond to a third measurement series, conducted after retracting the specimen holder to the loadlock position. For this measurement series, the voltage offset at $I_H$~=~0~mA after heating up is found to be higher than before, suggesting that the device underwent the initial process in reverse direction.

It is instructive to compare the measured I-V curves and voltage offsets with the measurements conducted on similar devices without gap \cite{Hettler2025}. The measured voltage offset in gap-free devices at $I_H$~=~0~mA was in the range of 10~$\upmu$V, thus being three orders of magnitude smaller than the offset measured for the nano-accordion. The thermo-voltage induced by the differential heating device followed a smooth, approximately quadratic evolution reaching maximum values of below 7~mV independent on the history of applied currents. This again strongly differs from the nano-accordion, which does not exhibit a well-defined dependency on $I_H$ and exhibits also considerably more noise. This comparison allows the conclusion, that although the induced thermo-voltage plays a role in the measured characteristics of the nano-accordion, it is clearly not the dominant contribution.

These considerations suggest that the electrical characteristics of the device cannot be explained solely by the (thermo)electric properties of the material but are related to the nano-gap, absent in all other studied devices. The observed unexpected electrical characteristics may be attributed to molecules in the gap region. Water and hydrocarbon molecules are commonly adsorbed on specimens under the high-vacuum conditions typical for TEM and can be assumed to be present on the inner surfaces of the gap region as well. As hydrocarbon molecules are nonpolar, it is difficult to imagine how these can generate an intrinsic voltage. Therefore, it is reasonable to assume that (polar) water molecules will have the largest impact on the electrical properties of the nano-accordion and cause the large voltage offset observed in the unheated state. Indeed, water molecules are known to have a significant impact on the work function of a material if chemisorbed on the surface and a monolayer of perfectly aligned molecules corresponds to a voltage step of 4~V \cite{Heras1980,Hettler2018}. The decrease of the voltage offset after heating under good vacuum conditions is then linked to the desorption of a considerable portion of the molecules due to the increasing temperature and facilitated by the mechanical movement of the device that causes the exposure of the interior surfaces of the gap and thus favors the desorption. In contrast, the retraction to the loadlock of the microscope, an area with worse vacuum conditions, leads to the re-adsorption of (water) molecules and thus to an increase of the observed voltage offset.

Figure~\ref{F:Figure4}c shows a HRTEM image of the gap region shortly after insertion in the  microscope. The image was acquired in slight overfocus conditions and the gap region appears with bright contrast compared to the vacuum region. Under overfocus conditions, atoms (positive potential) appear bright, \cite{Malac2017} indicating that indeed, material and/or positive charges are present in the gap region.

\subsection{Characterization after formation of a gap along the entire device}
\label{S:3.4}

\begin{figure}[t]
\centering
    \includegraphics[width=0.85\linewidth]{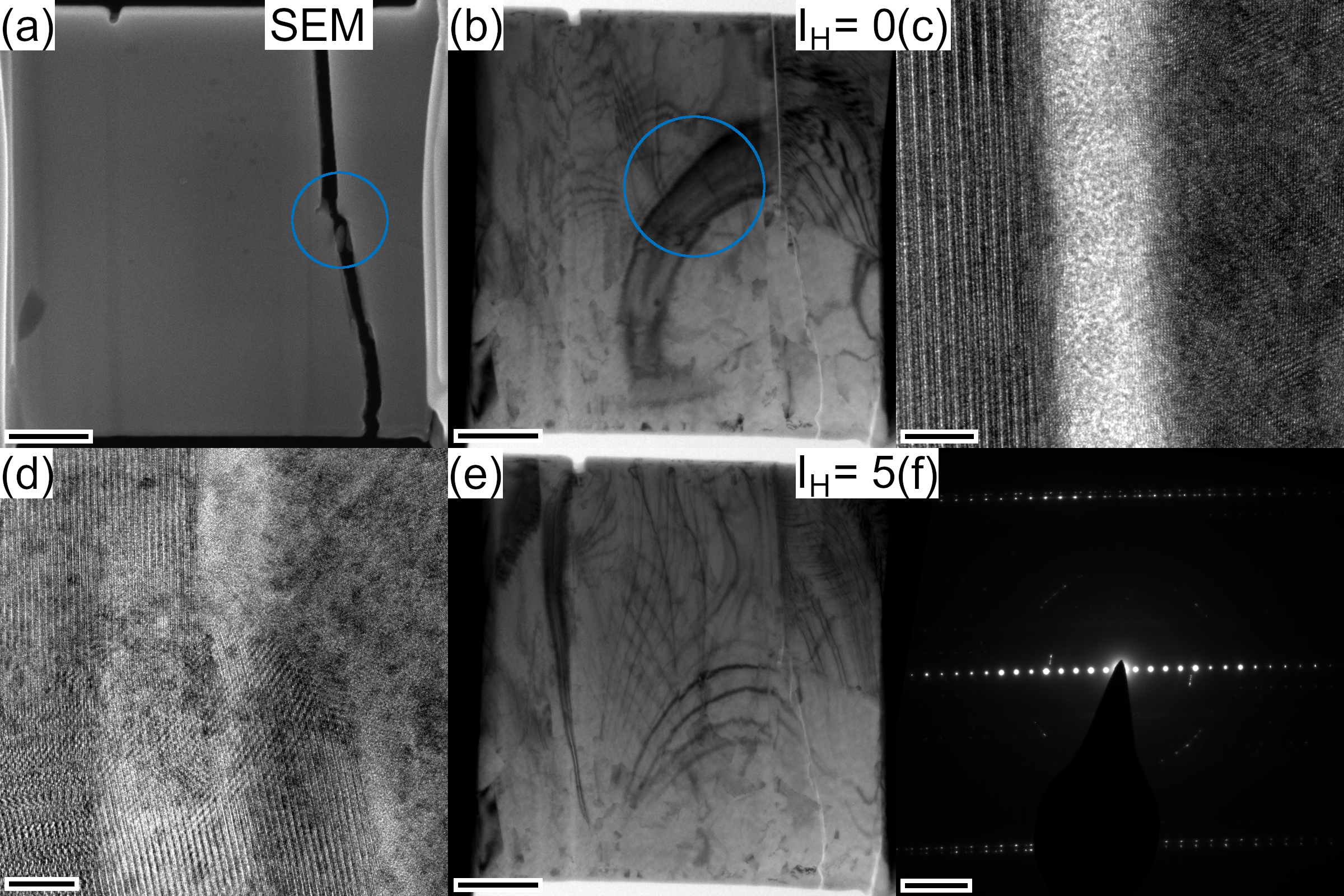}
    \caption{(a) SEM image of the device with full gap. The circle indicates the position with minimum distance. (b) BF-TEM images of the device with fully broken gap, circle indicates position for SAED pattern acquisition. (c) HRTEM image of the upper gap revealing amorphous material in the gap region. (d) HRTEM image of the triple-phase region showing that in this area, the gap only exhibits minimum width. (e) BF-TEM image of the device at an applied heating current of $I_H$~=~5~mA. (f) SAED pattern of the unheated device. Scale bars are (a,b,e) 300~nm, (c) 5~nm, (d) 10~nm, (f) 4~nm\textsuperscript{-1}.}
    \label{F:Figure5}
\end{figure}

After the conducted measurements presented in the preceding sections, the nano-accordion was mounted in the dual-beam machine with the aim to further thin the device and study the impact on the electrical characteristics. Figure~\ref{F:Figure5}a shows an SEM image of the device after mounting, which reveals that the gap opened across the entire lamella. The image reveals a gap width of about 50 nm. In the center of the gap (marked by a blue circle), the two parts of the lamella still seem to be in a minimum direct contact. A TEM analysis of the device after the full rupture (Figure~\ref{F:Figure5}b-d) reveals a reduced gap width of less than 10~nm. The gap is visible as a fine white line in the BF-TEM image shown in Figure~\ref{F:Figure5}b and amounts to approximately 8~nm in the upper part of the lamella as seen in Figure~\ref{F:Figure5}c, which  was taken from the initial gap region. The HRTEM image reveals the stacking periodicity of the MLC stack on the left side of the gap, while only weak fringes are visible on the right side. The gap is filled with an amorphous material, most probably from hydrocarbon molecules (contamination). The absence of a gap in the HRTEM image taken from the region of the triple-phase boundary (Figure~\ref{F:Figure5}d) suggests that, in that region, the two parts might still be in direct contact.

We address the difference in gap distance between the two microscopic techniques to electrostatic charging in SEM. The lower energy of the irradiating electrons results in the generation of a large number of secondary electrons, which typically cause positive charging of the insulating SiN membrane. This charging, combined with the relatively poor grounding provided by conductive tape could lead to an electrostatic repulsion of the two parts of the device and thus to the large gap. In fact, this electrostatic charging could have contributed to the full rupture of the device. In contrast, electrostatic charging in TEM is negligible and the grounding given by the contact pins of the specimen holder is better, leading to the apparent closing of the gap.

Interestingly, the mechanical response of the device with full gap to an applied differential heating current $I_H$ is highly similar when compared to the case of a partial gap. Supplementary video S5 shows the BF-TEM image evolution with increasing $I_H$ and reveals a highly similar bending contour contrast. Figure~\ref{F:Figure5}e shows a BF-TEM image of the unheated device, acquired under similar conditions compared to Figure~\ref{F:Figure2}e. A comparison of both images reveals that the bending contours are slightly broader in the left part of the device for the full gap, indicating a slightly reduced initial bending of the device. The movement of the bending contours then follows a similar evolution as observed for the partial gap as seen from supplementary video S5 and the BF-TEM acquired at $I_H$~=~5~mA shown in Figure~\ref{F:Figure5}e. The SAED pattern acquired in the unheated state (Figure~\ref{F:Figure5}f) again shows the [310] crystal orientation of the material.

\begin{figure}[t]
\centering
    \includegraphics[width=0.85\linewidth]{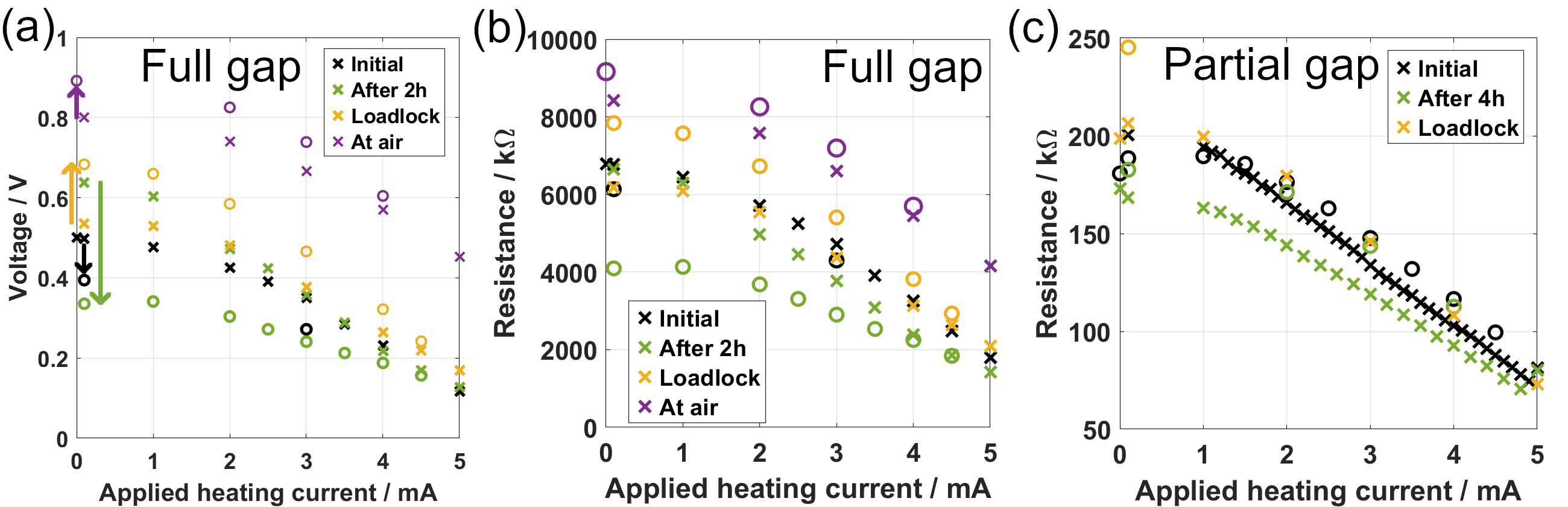}
    \caption{(a) Voltage offset plotted over the applied heating current for the device with full gap and for different conditions: Directly after insertion in the microscope (blue crosses/circles), after 2h in the microsocpe (red), after retraction to the loadlock (yellow) and at air (purple). Crosses represent the rise and circles the descent of the heating current. (b,c) Measured resistances for (b) the device with full gap and (c) the device with full gap, corresponding to the voltage offsets shown in (a) and in Figure~\ref{F:Figure5}d, respectively.}
    \label{F:Figure6}
\end{figure}

In contrast, a strong change is observed for the electrical properties. Figure~\ref{F:Figure6}a depicts the voltage offset determined in dependence of $I_H$, similarly as conducted for Figure~\ref{F:Figure4}b. The striking difference is the magnitude of the measured voltage offset, which is between 0.3 and 0.9 V, more than 30 times larger than for a partial gap. This confirms that the induced voltage offset is indeed related to the gap, as its contribution can be expected to be strongly increased in case of the full gap. Figure~\ref{F:Figure6}a shows the results for measurements conducted under different vacuum conditions. The black crosses and circles represent the first measurement after insertion into the microscope. Three additional measurements have been added: one measurement after 2h in the microscope (green symbols), a measurement after retraction to the loadlock (yellow symbols) and at ambient pressure after full retraction from the microscope (purple symbols). The shape is similar for all of the measurements, showing a general decrease of the induced voltage offset for increasing $I_H$. A difference is seen for the rise (crosses) and descent (circles) for all measurements, the direction however changes in dependence of the vacuum environment. Both measurements conducted in the high vacuum of the microscope column (black and green symbols) show a reduced voltage offset at $I_H$~=~0~mA after the current ramp, as indicated by an arrow in Figure~\ref{F:Figure6}a. In contrast, the voltage offset increases after application of the current in poor vacuum conditions as seen from the measurements performed in the loadlock and at air. 

When comparing the two measurements conducted in good vacuum, the voltage offset agrees for heating currents larger than 3~mA, reaching a value of 0.12~V at $I_H$~=~5~mA. However, a considerable difference is found in the voltage offset measured before performing the current ramp, which is 0.5~V for the initial measurement and 0.65~V for the measurement conducted after 2h. This is in contrast to the device with partial gap, where repeated measurements after the initial ramp resulted in similar curves, which was explained by the removal of adsorbed molecules from the gap region. This difference is attributed to the larger gap opening for the full gap. Firstly, a possible contribution could be minor mechanical changes of the device, which could lead to a decrease or increase of a possible direct contact between the left and right parts of the lamella after the application of a heating ramp. Secondly, the larger gap distance leads to a stronger exposure of the gap surfaces facilitating the adsorption of molecules from the vacuum environment even in a closed state. The molecules can therefore differ in both type and arrangement between the initial and the later measurement, leading to a change in the induced voltage offset in the unheated state. As hydrocarbon molecules are adsorbed with less adsorption energy, it is reasonable to assume that their presence and thus effect will reduce with the time of the device under high vacuum. The higher initial voltage offset after 2h in the microscope might thus be attributed to a higher amount of water molecules. In a heated state, the impact of the gap is reduced, leading to the agreement at higher applied currents. Similar to the partial gap, the heating and opening again leads to the removal of adsorbed molecules, which explains the reduced voltage offset after the heating cycle.

In addition to the voltage offset, the I-V curves also provide a value for the resistance of the device. Therefore, the curve is fitted by a linear function for lower applied voltages and the resistance is determined from the gradient. The resistances determined for both, full and partial gap are shown in dependence of the applied heating current in Figure~\ref{F:Figure6}b and c, respectively. The initial resistance in the unheated state strongly increases for the full gap by more than 30 times from approximately 200 k$\Upomega$ to 7~M$\Upomega$. This increase is expected as the device with partial gap still possessed a large continuous connection of the material, while such a direct connection is reduced to a minimum for the full gap. 
When neglecting a possible contribution of the gap to the resistance value of the device with partial gap, the resistance calculates to 0.6~$\Upomega$m, which is close to the literature value of the CCO device \cite{Torres.2022}, indicating that for the device with partial gap, the resistivity is governed by the material.

The measured resistance in the unheated state suggests that the partial gap only has a minor impact on the resistance, which is confirmed by the comparison of the resistance curves for the different vacuum conditions in Figure~\ref{F:Figure6}c. All curves show a highly similar trend towards lower resistance with increasing applied heating current, which can be attributed to the increase in conductivity of the semiconductive CCO material due to the increasing temperature. In contrast, the measured resistance strongly differs for the different vacuum environments and current histories in case of the full gap (Figure~\ref{F:Figure6}b). The differences between rise and descent are similar to the measured voltage offset. The resistance at $I_H$~=~0~mA decreases after the current ramp in case of good vacuum conditions (black and green symbols in Figure~\ref{F:Figure6}b) and otherwise increases after the current ramp (yellow and purple symbols). Again, these observations could be explained by the amount of molecules present in the gap region. Under good vacuum conditions, the contribution of poorly conducting water molecules are expected to decrease after heating up, possibly leading to a better contact between the two parts of the lamella. Under worse vacuum conditions, the trend is reversed. 

\section{Conclusion}

This work presents the nano-accordion, an \textit{in-situ} TEM device made of a misfit-layered compound material with implemented nano-gap and details its mechanical and electrical response to an applied differential heating current. The gap formed at a grained boundary and initially remained only partially spanning the device. TEM investigations for varying applied heating currents revealed a highly controllable mechanical bending of the device, which reminds the moving bellows of an accordion. The bending also induced the opening of the gap. The electrical characterizations show a high induced intrinsic voltage, which is attributed to (water) molecules adsorbed in the gap region, which are forced to desorb upon heating under high vacuum conditions, leading to a reduced intrinsic voltage in the unheated state. In contrast, the initial intrinsic voltage is restored when opening the gap under worse vacuum conditions, where (water) molecules can re-adsorb. The gap expanded to spread the whole device, which only had a small impact on the mechanical response to an applied heating current, but a strong influence on the electrical properties, which were found to be dominated by the full gap.

The work is of interest as it provides a way to prepare a nano-sized gap in a controlled way by specifically locating grain boundaries or pores in a material. Such electrically controlled gaps could be used to study, e.g., the electrical properties of specific gases. Moreover, this study allows drawing conclusions for the conduction of \textit{in-situ} TEM experiments as it reveals that considerable mechanical stresses and bending can occur upon \textit{in-situ} heating, which need to be taken into account for a correct interpretation of the acquired data.

\section*{Acknowledgments}

The authors acknowledge funding from the European Union’s Horizon 2020 research and innovation programme under the Marie Sklodowska-Curie grant agreement No 889546 and by the Spanish MICIU with funding from European Union Next Generation EU (PRTR-C17.I1) promoted by the Government of Aragon as well as from the Spanish MICIU (PID2023-151080NB-I00/AEI/10.13039/501100011033 and CEX2023-001286-S MICIU/AEI/10.13039/501100011033) and the Government of Aragon (DGA) through the project E13 23R. Sample courtesy from A. Sotelo (INMA, Universidad de Zaragoza) The microscopy works have been conducted in the Laboratorio de Microscopias Avanzadas (LMA) at Universidad de Zaragoza. 

\section*{Supplementary Information and videos}

Supplementary information file contains additional electron microscopy data.
\begin{itemize}
    \item Supplementary video 1: Evolution of BF-TEM images and measured voltage offset in dependence of applied heating current for partial gap. Scale bar is 300~nm.
    \item Supplementary video 2: Evolution of DF-TEM images in dependence of applied heating current for partial gap. Scale bar is 200~nm.
    \item Supplementary video 3: Evolution of HAADF-STEM images in dependence of applied heating current for partial gap. Scale bar is 100~nm.
    \item Supplementary video 4: Evolution of SAED patterns in dependence of applied heating current for partial gap. Scale bar is 3~nm\textsuperscript{-1}.
    \item Supplementary video 5: Evolution of BF-TEM images and measured voltage offset in dependence of applied heating current for full gap. Scale bar is 300~nm.
\end{itemize}

\section*{Data availability}
Data is available under https://doi.org/10.5281/zenodo.17549328.

\bibliographystyle{citstyle.bst}
\bibliography{Biblio}

@article{Torres.2022,
 author = {Torres, M. A. and Madre, M. A. and Dura, O. J. and Garc{\'i}a, G. and Marinel, S. and Martinez-Filgueira, P. and Sotelo, A.},
 year = {2022},
 title = {Evaluation of pressure and temperature effect on the structure and properties of Ca2.93Sr0.07Co4O9 ceramic materials},
 pages = {7730--7747},
 volume = {48},
 number = {6},
 issn = {02728842},
 journal = {Ceramics International},
 doi = {10.1016/j.ceramint.2021.11.321}
}

@Article{Hettler2025,
  author    = {Hettler, Simon and Furqan, Mohammad and Sotelo, Andrés and Arenal, Raul},
  journal   = {Ultramicroscopy},
  title     = {Toward quantitative thermoelectric characterization of (nano)materials by in-situ transmission electron microscopy},
  year      = {2025},
  issn      = {0304-3991},
  month     = jan,
  pages     = {114071},
  volume    = {268},
  doi       = {10.1016/j.ultramic.2024.114071},
  publisher = {Elsevier BV},
}

@Article{Miyazaki2002,
  author    = {Miyazaki, Yuzuru and Onoda, Mitsuko and Oku, Takeo and Kikuchi, Masae and Ishii, Yoshinobu and Ono, Yasuhiro and Morii, Yukio and Kajitani, Tsuyoshi},
  journal   = {Journal of the Physical Society of Japan},
  title     = {Modulated Structure of the Thermoelectric Compound [Ca2CoO3]0.62CoO2},
  year      = {2002},
  issn      = {1347-4073},
  month     = feb,
  number    = {2},
  pages     = {491--497},
  volume    = {71},
  doi       = {10.1143/jpsj.71.491},
  publisher = {Physical Society of Japan},
}

@Article{Seto2022,
  author    = {Seto, Yusuke and Ohtsuka, Masahiro},
  journal   = {Journal of Applied Crystallography},
  title     = {ReciPro: free and open-source multipurpose crystallographic software integrating a crystal model database and viewer, diffraction and microscopy simulators, and diffraction data analysis tools},
  issn      = {1600-5767},
  year      = {2022},
  pages     = {397--410},
  volume    = {55},
  doi       = {/10.1107/S1600576722000139},
}

@article{Mayadas1970,
  title = {Electrical-Resistivity Model for Polycrystalline Films: the Case of Arbitrary Reflection at External Surfaces},
  author = {Mayadas, A. F. and Shatzkes, M.},
  journal = {Phys. Rev. B},
  volume = {1},
  issue = {4},
  pages = {1382--1389},
  numpages = {0},
  year = {1970},
  month = {Feb},
  publisher = {American Physical Society},
  doi = {10.1103/PhysRevB.1.1382},
}

@article{Andrews1969,
author = {P. V. Andrews and M. B. West and C. R. Robeson},
title = {The effect of grain boundaries on the electrical resistivity of polycrystalline copper and aluminium},
journal = {The Philosophical Magazine: A Journal of Theoretical Experimental and Applied Physics},
volume = {19},
number = {161},
pages = {887--898},
year = {1969},
publisher = {Taylor \& Francis},
doi = {10.1080/14786436908225855}
}

@article{Hettler2018,
title = {Charging of carbon thin films in scanning and phase-plate transmission electron microscopy},
journal = {Ultramicroscopy},
volume = {184},
pages = {252-266},
year = {2018},
issn = {0304-3991},
doi = {https://doi.org/10.1016/j.ultramic.2017.09.009},
url = {https://www.sciencedirect.com/science/article/pii/S0304399117303169},
author = {Simon Hettler and Emi Kano and Manuel Dries and Dagmar Gerthsen and Lukas Pfaffmann and Michael Bruns and Marco Beleggia and Marek Malac}
}

@article{Heras1980,
title = {Work function changes upon water contamination of metal surfaces},
journal = {Applications of Surface Science},
volume = {4},
number = {2},
pages = {238-241},
year = {1980},
issn = {0378-5963},
doi = {https://doi.org/10.1016/0378-5963(80)90133-6},
url = {https://www.sciencedirect.com/science/article/pii/0378596380901336},
author = {J.M. Heras and L. Viscido}
}

@article{Malac2017,
title = {Computer simulations analysis for determining the polarity of charge generated by high energy electron irradiation of a thin film},
journal = {Micron},
volume = {100},
pages = {10-22},
year = {2017},
issn = {0968-4328},
doi = {https://doi.org/10.1016/j.micron.2017.03.015},
url = {https://www.sciencedirect.com/science/article/pii/S0968432816303870},
author = {Marek Malac and Simon Hettler and Misa Hayashida and Masahiro Kawasaki and Yuji Konyuba and Yoshi Okura and Hirofumi Iijima and Isamu Ishikawa and Marco Beleggia}
}

@article{BUENOVILLORO2023118816,
title = {Fe segregation as a tool to enhance electrical conductivity of grain boundaries in Ti(Co,Fe)Sb half Heusler thermoelectrics},
journal = {Acta Materialia},
volume = {249},
pages = {118816},
year = {2023},
issn = {1359-6454},
doi = {https://doi.org/10.1016/j.actamat.2023.118816},
author = {Ruben {Bueno Villoro} and Maxwell Wood and Ting Luo and Hanna Bishara and Lamya Abdellaoui and Duncan Zavanelli and Baptiste Gault and Gerald Jeffrey Snyder and Christina Scheu and Siyuan Zhang}
}

@article{MYPATI2022100223,
title = {TEM analysis and molecular dynamics simulation of graphene coated Al-Cu micro joints},
journal = {Carbon Trends},
volume = {9},
pages = {100223},
year = {2022},
issn = {2667-0569},
doi = {https://doi.org/10.1016/j.cartre.2022.100223},
author = {Omkar Mypati and Polkampally Pavan Kumar and Surjya Kanta Pal and Prakash Srirangam}
}

@article{DONG20231459,
title = {Contribution of grain boundary to strength and electrical conductivity of annealed copper wires},
journal = {Journal of Materials Research and Technology},
volume = {26},
pages = {1459-1468},
year = {2023},
issn = {2238-7854},
doi = {https://doi.org/10.1016/j.jmrt.2023.08.012},
author = {Liming Dong and Fei Yang and Tianbo Yu and Ning Zhang and Xuefeng Zhou and Zonghan Xie and Feng Fang}
}

@article{Graham2010,
    author = {Graham, R. L. and Alers, G. B. and Mountsier, T. and Shamma, N. and Dhuey, S. and Cabrini, S. and Geiss, R. H. and Read, D. T. and Peddeti, S.},
    title = {Resistivity dominated by surface scattering in sub-50 nm Cu wires},
    journal = {Applied Physics Letters},
    volume = {96},
    number = {4},
    pages = {042116},
    year = {2010},
    month = {01},
    issn = {0003-6951},
    doi = {10.1063/1.3292022}
}

@article{Bishara2021,
author = {Bishara, Hanna and Lee, Subin and Brink, Tobias and Ghidelli, Matteo and Dehm, Gerhard},
title = {Understanding Grain Boundary Electrical Resistivity in Cu: The Effect of Boundary Structure},
journal = {ACS Nano},
volume = {15},
number = {10},
pages = {16607-16615},
year = {2021},
doi = {10.1021/acsnano.1c06367}
}

@article{Aslam2011,
author = {Aslam, Zabeada and Nicholls, Rebecca and A. Koos, Antal and Nicolosi, Valeria and Grobert, Nicole},
title = {Investigating the Structural, Electronic, and Chemical Evolution of B-Doped Multi-walled Carbon Nanotubes as a Result of Joule Heating},
journal = {The Journal of Physical Chemistry C},
volume = {115},
number = {50},
pages = {25019-25022},
year = {2011},
doi = {10.1021/jp206424v}
}

@article{Hettler_2021,
doi = {10.1088/2053-1583/abedc9},
url = {https://dx.doi.org/10.1088/2053-1583/abedc9},
year = {2021},
month = {apr},
publisher = {IOP Publishing},
volume = {8},
number = {3},
pages = {031001},
author = {Hettler, Simon and Sebastian, David and Pelaez-Fernandez, Mario and Benito, Ana M and Maser, Wolfgang K and Arenal, Raul},
title = {In-situ reduction by Joule heating and measurement of electrical conductivity of graphene oxide in a transmission electron microscope},
journal = {2D Materials}
}

@article{Hsueh2023,
author = {Hsueh, Yu-Hsiang and Ranjan, Ashok and Lyu, Lian-Ming and Hsiao, Kai-Yuan and Chang, Yu-Cheng and Lu, Ming-Pei and Lu, Ming-Yen},
title = {In Situ/Operando Studies for Reduced Eletromigration in Ag Nanowires with Stacking Faults},
journal = {Advanced Electronic Materials},
volume = {9},
number = {3},
pages = {2201054},
doi = {https://doi.org/10.1002/aelm.202201054},
year = {2023}
}

@article{Hettler2024,
author = {Hettler, Simon and Sreedhara, MB and Tenne, Reshef and Arenal, Raul},
title = {Unraveling the Decomposition Pathways of LaS-TaS2 Misfit-Layered Compound Nanostructures under Extreme Electrical Currents by In Situ TEM},
journal = {The Journal of Physical Chemistry C},
volume = {129},
number = {30},
pages = {13803-13812},
year = {2025},
doi = {10.1021/acs.jpcc.5c03498}
}

@article{Ng2022,
    author = {Ng, Nicholas and McQueen, Tyrel M.},
    title = {Misfit layered compounds: Unique, tunable heterostructured materials with untapped properties},
    journal = {APL Materials},
    volume = {10},
    number = {10},
    pages = {100901},
    year = {2022},
    month = {10},
    issn = {2166-532X},
    doi = {10.1063/5.0101429}
}

@article{Fava2021,
  title = {Effects of doping substitutions on the thermal conductivity of half-Heusler compounds},
  author = {Fava, Mauro and Dongre, Bonny and Carrete, Jes\'us and van Roekeghem, Ambroise and Madsen, Georg K. H. and Mingo, Natalio},
  journal = {Phys. Rev. B},
  volume = {103},
  issue = {17},
  pages = {174112},
  numpages = {10},
  year = {2021},
  month = {May},
  publisher = {American Physical Society},
  doi = {10.1103/PhysRevB.103.174112},
  url = {https://link.aps.org/doi/10.1103/PhysRevB.103.174112}
}

\end{document}